# Enhanced Frequency Noise Discrimination Using Cavity-coupled Mach-Zehnder Interferometer

**Mohamad Hossein Idjadi[1,*] and Andrea Blanco-Redondo[1]**

[1]Bell Laboratories, Nokia, 600 Mountain Avenue, Murray Hill, New Jersey 07974, USA
*corresponding author: mohamad.idjadi@nokia-bell-labs.com

**Here, an enhanced frequency noise discriminator is proposed using a cavity-coupled Mach-Zehnder interferometer and demonstrated on a general-purpose programmable integrated photonics platform. The frequency noise measurement sensitivity is similar to the well-known Pound-Drever-Hall architecture but owns a passive, simpler, and more compact design. The proposed structure can be utilized in integrated laser frequency stabilization architectures.**

**INTRODUCTION**

Frequency noise discriminators play an essential role in laser frequency noise characterization and frequency/phase stabilization. Different frequency noise discriminator architectures have been proposed, including a Fabry-Perot cavity in a Pound-Drever-Hall (PDH) laser frequency-locking architecture [1] and an unbalanced Mach-Zehnder interferometer (MZI) [2-6]. The PDH control loop, despite its excellent performance, requires fast optical phase modulation that leaves residual amplitude noise and ultimately degrades the laser noise performance [7,8]. On the other hand, an unbalanced MZI can be used to detect laser frequency noise without a need for phase modulation. However, the unbalanced MZIs proposed to date require large space on-chip for similar discrimination sensitivity. The cavity-coupled MZI structure with a large number of cascaded ring resonators in the arms of the MZI has been proposed as an optical filter to improve the linearity of frequency discrimination in broad-band microwave-photonic links [6]. Here, we propose, analyze, and demonstrate the use of a cavity-coupled MZI as a laser frequency noise discriminator that offers high sensitivity in noise measurement while can potentially occupy a fraction of the photonic chip area compared to its counterparts, thanks to the resonance effect. The proposed architecture provides sensitive frequency noise measurement capability that can be utilized in an integrated electro-optic feedback loop to detect laser frequency fluctuations, process the error signal in the electrical domain, and feed it back to the laser to suppress its frequency noise.

**PRINCIPLE OF OPERATION**

Figure 1(a) shows a generalized block diagram of a frequency discriminator where it receives an incoming electric field oscillating at the optical frequency ($\omega_0$), compares its frequency fluctuations with a frequency reference ($\omega_{ref}$), and generates a proportional electronic signal. This electrical signal, *i.e.,* the error signal, can be used to either characterize the frequency noise of a laser (Fig. 1(b)) or to stabilize its frequency when used in a feedback control loop (Fig. 1(c)) [4]. It is worth mentioning that the performance of a frequency discriminator can be quantified as its sensitivity to frequency fluctuations (*i.e.,* the slope of the current-frequency curve). Figure 2 shows three examples of frequency discriminator architectures with optical input and electrical output ports: the PDH loop, an MZI with true-time delay, and the proposed cavity-coupled MZI. In Fig. 2(b) and (c), the error signal ($I_{out}$) at the output of the balanced photodetectors can be written as

$$I_{out}(\omega) = RP_0|T(\omega)|\sin(\sphericalangle T(\omega) - \phi), \qquad (1)$$

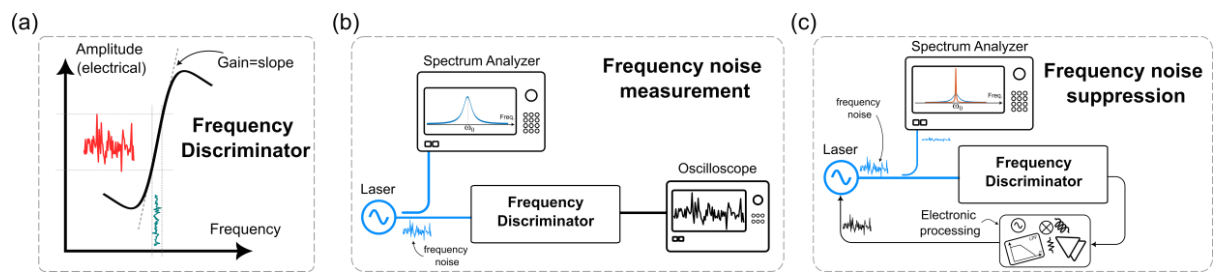

*Fig. 1. (a) The frequency noise discriminator. Potential applications of a frequency noise discriminator in (b) laser frequency noise measurement (b) and laser frequency noise suppression using an electro-optic feedback loop.*

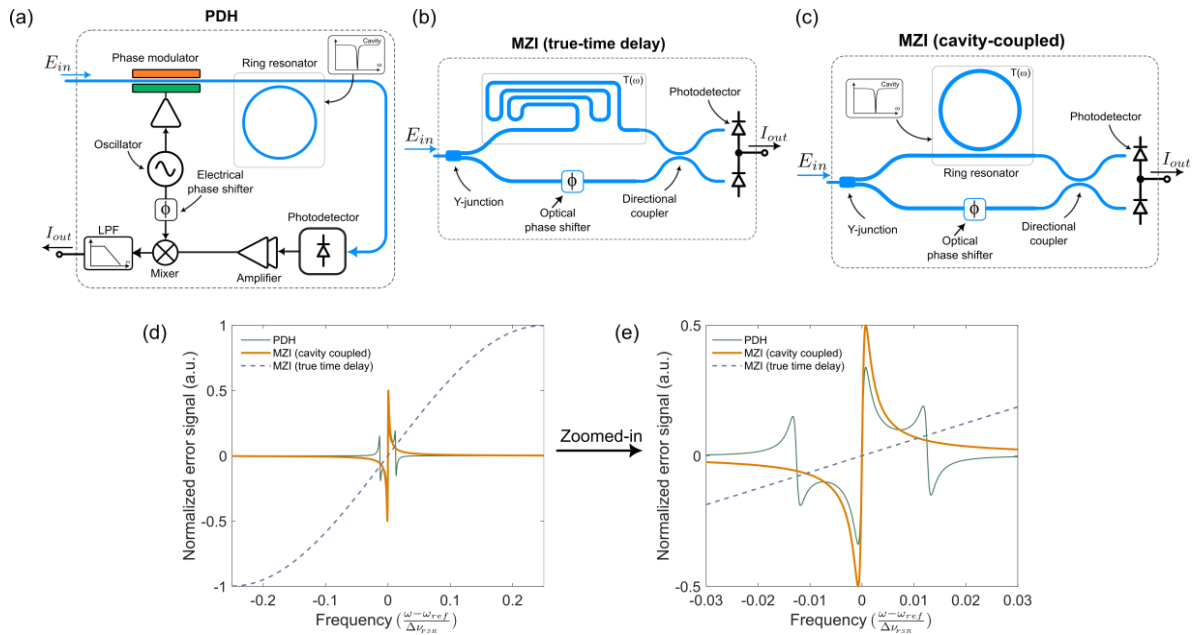

Fig. 2. (a) The PDH frequency discriminator. A laser is phase modulated using a local oscillator (LO) and is filtered by a ring resonator. The ring resonator acts as a frequency reference. The output of the ring resonator is photo-detected, amplified, down-converted with the same LO frequency, and finally, low-pass filtered. (b) A conventional MZI with a true-time delay in one arm. An optical phase shifter is used to adjust the phase of the two arms of the MZI. The MZI is terminated with balanced photodetectors. (c) The cavity-coupled MZI. In this architecture, a ring resonator is used as a phase and frequency reference, and its output is coherently detected using balanced photodetectors. (d, e) Simulation of the normalized error signals.

where $R, P_0, |T(\omega)|, \angle T(\omega), \omega$, and $\phi$ are photodetector responsivity, laser power, amplitude, and phase of the reference transfer function, laser instantaneous frequency, and the static bias phase on the other arm of the MZI, respectively. Without loss of generality, we can consider a high quality-factor (Q-factor) ring resonator or a low loss waveguide as the frequency reference and the error signal ($I_{out}$) can be calculated numerically using Eq. (1). For the numerical calculation, the waveguide loss is assumed 0.2 dB/cm and the ring circumference in the case of PDH loop and the cavity-coupled MZI are identical and equal to the length difference in an unbalanced conventional MZI architecture in Fig. 2(b) (L= 1 mm). This corresponds to a Q-factor of about 1.6 million for the ring resonator at the resonance wavelength of 1550 nm. Also, the local oscillator frequency in PDH architecture is 1 GHz, and the optical and electrical phase shifters are optimized accordingly. Figure 2(d) shows the comparison between the normalized error signal for these three cases. As shown in Fig. 2(d), all three error signals are asymmetric around the reference frequency ($\omega_{ref}$), which is the resonance frequency of the ring resonator and the quadrature point of the MZI with true-time delay. This asymmetric response is essential for the laser frequency control loop to distinguish the frequency compensation direction required for frequency locking. As shown in Fig. 2(e), it is clear that given the same waveguide length and electronics gain, the PDH and the cavity-coupled MZI (Figs. 2(a) and 2(c)) have considerably higher frequency detection gain (i.e., the slope at $\omega = \omega_{ref}$) compared to a conventional MZI (Fig. 2(b)). The frequency discrimination gain for PDH, MZI with true-time delay, and cavity-coupled MZI are 1.14×10$^{-8}$ [Hz$^{-1}$], 8×10$^{-11}$ [Hz$^{-1}$], and 1.7×10$^{-8}$ [Hz$^{-1}$], respectively. In general, the higher the Q-factor of the ring, the higher the noise discrimination sensitivity. Thanks to the resonance effect in the cavity-couple MZI, the frequency discrimination sensitivity is more than two orders of magnitude higher than a conventional MZI but unlike the PDH architecture, no phase modulation and complex electronics are required.

**EXPERIMENTAL DEMONSTRATION**

Figure 3 shows the proof-of-concept experimental demonstration of the proposed cavity-coupled MZI frequency noise discriminator. We implemented the proposed photonic integrated circuit (PIC) on a general-purpose programmable photonics platform consisting of a programmable mesh of integrated 2x2 directional couplers and phase shifters (iPronics SmartLight Processor) [9]. Figure 3(a) shows the implemented PIC in which different photonic functionalities (e.g. power splitter, ring resonator, and MZI) can be implemented by adjusting the coupling ratio of each programmable unit cell (PUCs), *i.e.,* tuneable couplers. Figure 3(b) shows the equivalent schematic for the implemented PIC where the input light ($E_{in}$) is coupled to the structure and the ring resonator output ($E_{out0}$) and the MZI outputs ($E_{out1(+)}, E_{out1(-)}$) are coupled out using optical fibers. The phase of the MZI is adjusted using thermal phase shifters. Figure 3(c) shows the measurement setup. A tuneable 10 mW continuous wave laser is coupled into the programmable PIC and its wavelength is swept from 1549 nm to 1551 nm with 1 pm resolution. The PUC settings are uploaded by a computer and the output electric fields, $E_{out1(+)}$, $E_{out1(-)}$, and $E_{out0}$, are photo-detected using Thorlabs PDB470C and PDB450C photodetectors and monitored on an oscilloscope. Figure 3(d) shows the normalized resonance response of the ring resonator coupled to the MZI. Given the loss of each PUC (~0.48 dB), the measured ring resonator Q-factor and the extinction ratio are about 73K and 12 dB, respectively. The PUC basic unit length is 811.41 μm which sets the ring FSR to about 15 GHz. Figure 3(e) shows the measured error signal which agrees well with the mathematical model. As shown in Fig. 3(e), the error signal is asymmetric around the reference ring resonance frequency with the normalized frequency noise discriminator gain of about $10^{-9}$ [Hz$^{-1}$]. This gain is limited by the PUC's insertion losses and consequently the limited ring resonator Q-factor.

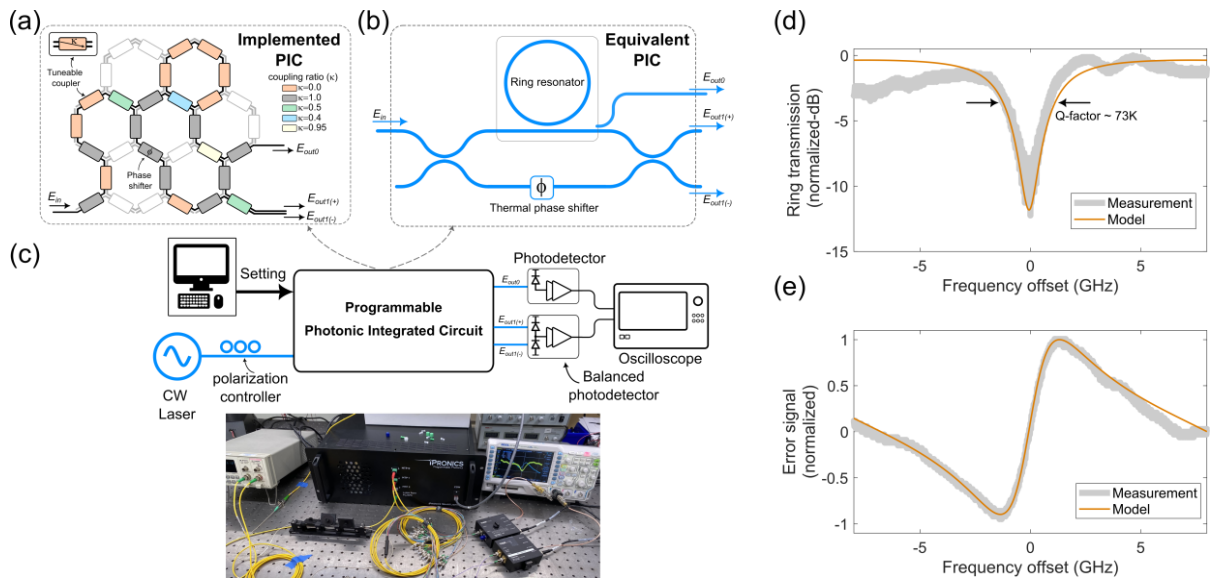

*Fig. 3. Experimental demonstration of the proposed cavity-coupled MZI acting as a frequency noise discriminator. (a) The implemented PIC is based on a programmable mesh of integrated directional couplers. (b) The equivalent schematic of the implemented PIC. (c) the measurement setup. (d) The normalized ring resonator response, and (e) the measured error signal. The frequency offset is defined relative to the ring resonance frequency.*

**DISCUSSION**

Here we propose, analyze, and experimentally demonstrate the use of a cavity-coupled MZI as an enhanced frequency noise discriminator. The proposed architecture leverages the resonance effect of the cavity and coherent detection in an MZI architecture. The frequency noise detection sensitivity of the proposed architecture is similar to the well-known PDH architecture while it owns a less complex design, lower power consumption control electronics, and does not require phase modulation which would, inevitably, introduce residual amplitude modulation. Finally, as a proof of concept, we implemented this architecture on an integrated programmable photonic platform. The limitations of the programmable photonic platform can be solved by implementing the PIC on an application-specific tape-out process. The enhanced frequency noise discrimination can be utilized in a variety of applications such as laser frequency noise suppression to achieve compact low-cost narrow linewidth lasers.